\documentstyle[preprint,aps]{revtex}
\def\hb{\bar{H}}
\def\bs{\bar{\psi}}
\begin{document}
\title{ON AN EXPLICIT  SET  OF  COMPLETE  EIGENFUNCTIONS  FOR  THE 
CALOGERO-SUTHERLAND MODEL}
\author{N. Gurappa   and   Prasanta. K. Panigrahi$^\dagger$}
\address{School of Physics\\University of Hyderabad\\
Hyderabad - 500 046, {\bf INDIA}}
\maketitle
\begin{abstract}
An operator method is provided to generate an explicit set of complete 
eigenfunctions for the Calogero-Sutherland model, obtained earlier by Vacek, 
Okiji and Kawakami, through a special ansatz. We find the connection of the
above basis set with the general eigenfunctions of this model and explicitly
show that these states describe {\it only} the center of mass motion.
\end{abstract}
\vfill
$^\dagger$email: panisp@uohyd.ernet.in
\newpage

Calogero-Sutherland (CS) model,$^1$ and its generalizations to spin-chains,$^2$
have attracted wide attention in the recent literature because of their 
relevance to many branches of physics.$^3$ This quantum mechanical model in 
one dimension represents a system of $N$-identical particles, simultaneously 
subjected to a harmonic confinement and a pairwise inverse square potential. 
The correlated ground state of this model was first analysed by Sutherland$^1$ 
in the context of thermodynamics, who also pointed out its connection with the 
random matrices.$^4$ The algebraic structure of the CS model, studied earlier 
by Perelomov through group theoretical means,$^5$ has recently been further 
investigated by a number of authors, using the $S_N$-{\it extended} Heisenberg 
algebra.$^6$ Much akin to the harmonic oscillator case, the excited state 
wavefunctions are constructed by operating appropriately symmetrized monomials 
of the raising operators on the vacuum, which is annihilated by the lowering 
operators. However, in practice, the construction of an explicit basis set has 
not proceeded much further than a few particle case,$^7$ due to the complicated
sums appearing in the definition of the creation and annihilation operators.

Recently, a special set of complete eigenfunctions spanning {\it all} energy 
levels of the CS Hamiltonian  $( \hbar = \omega=m=1)$
\begin{equation}
H  =  -{1\over2} \sum_i^N \partial_i^2 +  {1\over2}  \sum_i^N  x_i^2  + 
{g^2\over2} \sum_{{i,j} \atop {i \ne j}}^N {1\over(x_i-x_j)^2} \qquad ,
\end{equation}
was constructed by Vacek, Okiji and Kawakami (VOK)$^8$
through a Jastrow-type ansatz. The states are characterized by an integer $I$:
\begin{equation}
\psi_I  =   \left[\prod_{i<j}^N{|x_i-x_j|^\lambda} {(x_i - x_j)^\gamma} \right]
\exp\left[-{1\over2}\sum_i^N x_i^2\right] \sum_{\sum_im_i=I} \prod_i^N 
{H_{m_i}(x_i)\over m_i!} \qquad ,
\end{equation}
with an eigenvalue  $E_I=E_o+I$.
Here, $\gamma = 0$ $(\gamma = 1)$ and $g^2 = (\lambda + \gamma)
(\lambda + \gamma - 1)$ for the bosonic (fermionic) case and 
$E_o ={N\over2} + {1\over2} (\lambda + \gamma ) N(N-1)$ is the ground state
energy which depends on the interaction parameter $\lambda$.

A thorough analysis of these eigenfunctions is of relevance, since a 
basis set is crucial for the computation of the correlation functions (CF). In 
the case of CS model on a circle, the CF are calculated by using the properties
of Jack polynomials.$^9$ Furthermore, the basis functions may clarify various
intriguing aspects of this model, one of which is the occurrence of
quasi-particles with the so called exclusion statistics.$^{10}$

In this paper, we first provide an operator method based on the
$S_N$-{\it extended} Heisenberg algebra, to generate the above mentioned 
eigenstates. This is done by introducing suitable raising and lowering
operators, keeping in mind the identical nature of the particles and the 
symmetry of the wavefuctions under transposition of the particle coordinates. 
Powers of this creation operator acting on the ground state produce the 
desired states given by VOK. It is then found, using the variable separation
method, that these eigenfunctions are a special, albeit complete subset of the 
most general wavefunctions of the CS model. Furthermore, it is shown that 
these states describe {\it only} the center of mass motion. 

The $S_N$-{\it extended} Heisenberg algebra is generated by the operators
$a_i$, $a_i^\dagger$ and $K_{ij}$ which satisfy the following relations 
\begin{eqnarray}
{}[a_i, a_j] &  = &  [a^\dagger_i, a^\dagger_j] = 0 \,\,\,,\nonumber\\
{}[a_i, a_j^\dagger] & = &  \delta_{ij}\{1+(\lambda + \gamma )  
\sum_l  K_{il}\} - (\lambda + \gamma ) K_{ij} \,\,\,,\nonumber\\
{}K_{ij} & =&  K_{ji}\,\,\,;\qquad  (K_{ij})^2 = 1 \,\,\,,\nonumber\\
{}K_{ij}a_j &  = & a_iK_{ij}\,\,\,;\qquad K_{ij}a^\dagger_j    
= a^\dagger_iK_{ij} \qquad  \mbox{(no summation over repeated indices)}, 
\nonumber\\
{}K_{ij}K_{jl}  & =&  K_{jl}K_{il}  =    K_{il}K_{ij}  \,\,\,,
\qquad  \mbox{for} \,\,\,i\ne j,\,\, i\ne l,\,\, j\ne l \,\,\,,\nonumber\\
{}K_{ij}K_{mn} & =&  K_{mn} K_{ij} \,\,\,,\,\,\ \mbox{for} \,\,\,  i,j,m,n 
\,\,\, \mbox{all different}.
\end{eqnarray}
It can be checked, after a straightforward calculation, that the 
$S_N$-{\it extended} Heisenberg algebra follows with, 
$a_i  =  {1\over\sqrt{2}} (x_i+D_i) $ 
and $a_i^\dagger  =  {1\over\sqrt{2}}(x_i-D_i)$.
Here,
\begin{equation}
D_i  = \partial_i+(\lambda + \gamma ) \sum_{j\atop{j \ne i}}^N 
{1\over(x_i-x_j)} (1-K_{ij}) 
\end{equation}
is the so called Dunkl derivative.$^{11}$

Based on this algebra, one can construct a Hamiltonian $\hb$, whose connection 
with the CS model will be established shortly, in the form given below,

\newpage
\begin{eqnarray}
\hb  =  {1\over2} \sum_i^N \{a_i, a_i^\dagger\}
& = &\left(-{1\over2}\sum_i^N \partial_i^2 + {1\over2} \sum_i^N x_i^2  
-(\lambda + \gamma ) \sum_{{i,j}\atop{i\ne j}}^N  {1\over(x_i-x_j)}  
\partial_i \right) \nonumber\\
& + &  \frac {(\lambda + \gamma )}{2} \sum_{{i,j}\atop{i\ne j}}^N 
{1\over(x_i-x_j)^2} (1-K_{ij})\nonumber\\
& - & \frac {(\lambda + \gamma )^2}{2}\sum_{{i,j,l}\atop{i\ne {j , l}}}^N
{1\over(x_i-x_j)} (1-K_{ij}) {1\over(x_i-x_l)} (1-K_{il}) \qquad .
\end{eqnarray}
Making use of Eq. (3), the following commutation relationships (CR) can be 
worked out,
\begin{eqnarray}
{}[\hb, a_i] & = & -a_i \qquad,\nonumber\\
{}[\hb, a^\dagger_i] & = & a^\dagger_i \qquad.
\end{eqnarray}
A generic excited eigenstate $\bs$ for  $\hb$, can then be written as 
\begin{equation}
\bs = \prod_i^N (a^\dagger_i)^{m_i} \psi_o \qquad .
\end{equation}
Here, $\psi_o$ is the ground state wavefunction obtained from 
\begin{equation}
a_i\psi_o = 0 \qquad.
\end{equation}

An eigenstate of $\hb$, satisfying $K_{ij}\bs=\bs$, {\it i.e.}, symmetric
under the transposition of the particle coordinates, is also an eigenstate of 
another related Hamiltonian $H^\prime$, where 
\begin{equation}
H^\prime   =  \left( -{1\over2} \sum_i^N  \partial_i^2  +  {1\over2}  
\sum_i^N x_i^2 - (\lambda + \gamma )  \sum_{{i,j}\atop{i\ne j}}^N  {1\over(x_i-x_j)} 
 \partial_i \right) \qquad.\\
\end{equation}
$H^\prime$ and the CS Hamiltonian $H$ in Eq. (1) are related by a similarity 
transformation 
of the form 
\begin{equation}
H^\prime  =  Z^{-(\lambda + \gamma)}\,\ H \,\ Z^{(\lambda + \gamma)} \qquad ,
\end{equation}
where,
$$Z =  \prod_{i<j}^N (x_i-x_j) \qquad .$$

Noting the fact that, the eigenstates of $H^\prime$ are symmetric under the
exchange of particle coordinates, we introduce the raising and lowering 
operators, respectively given by
\begin{eqnarray}
A  =  {1\over\sqrt{N}} \sum_i^N a_i & = & {1\over\sqrt{2N}} 
\left(\sum_i^N(x_i+\partial_i)  + (\lambda + \gamma )
\sum_{{i,j}\atop{i\ne j}}^N {1\over(x_i-x_j)} (1-K_{ij})\right) \nonumber \\
& = & {1\over\sqrt{2N}} \left(\sum_i^N(x_i+\partial_i)\right)  
\end{eqnarray}
and
\begin{eqnarray}
A^\dagger  =  {1\over\sqrt{N}} \sum_i^N a^\dagger_i ={1\over\sqrt{2N}} 
\left(\sum_i^N(x_i-\partial_i)\right) \qquad .
\end{eqnarray}
It is straightforward to check that $[A, A^\dagger]   = 1 $ and the CR of 
$A$ and $A^\dagger$ with $\hb$ are identical to the ones given in Eq. (6).

It is worth pointing out that, although $A$ and $A^{\dagger}$ are identical to
the lowering and raising operators of the harmonic oscillator algebra and have
similar CR with $\hb$, the Hamiltonian in the present case
can not be written as a bilinear of these operators. 

Defining the ground state $\phi_o$ as
$$A\phi_o =  {1\over\sqrt{2N}} \sum_i^N(x_i+\partial_i) \phi_o = 0 \qquad ,$$
one finds,
\begin{equation}
\phi_o = \exp\left(-{1\over2} \sum_i^N x_i^2\right) \qquad.
\end{equation}
Now, the excited states $\phi_I$'s, can be obtained by the repeated 
application of $A^\dagger$ on $\phi_o$.
The $I$-th excited state is
\begin{eqnarray}
\phi_I  =  (A^\dagger)^I\phi_o  & = &  {I!\over(N)^{I/2}}   \sum_{\sum_im_i=I} 
 \prod_i^N {(a^\dagger_i)^{m_i}\over m_i} \phi_o \nonumber\\
& = & {I!\over(N)^{I/2}} \exp\left[-{1\over2}\sum_i ^N x_i^2\right] 
\sum_{\sum_i m_i=I} \prod_i^N {H_{m_i}(x_i)\over m_i!} \qquad.
\end{eqnarray}

These eigenstates are obviously symmetric under the exchange of 
particle coordinates and hence we conclude that the $\phi_I$'s are also the 
eigenfunctions of $H^\prime$. An eigenfunction $\psi_I$ of the CS Hamiltonian 
can then be written as,
\begin{equation}
\psi_I  =  {I!\over(N)^{I/2}}  \left[\prod_{i<j}^N|x_i-x_j|^\lambda  
(x_i - x_j)^ \gamma
\right] \exp\left[-{1\over2}\sum_i^N x_i^2\right] \sum_{\sum_im_i=I} \prod_i^N 
{H_{m_i}(x_i)\over m_i!} \qquad .
\end{equation}

These eigenfunctions, modulo an overall normalization factor, are identical to
the one given in Eq. (2), obtained earlier by VOK, through a special ansatz. It
should be noted that, unlike the group theoretical approach, this method does 
not account for the degeneracy of the energy eigenvalues.

Now, we show that the above set is a special, albeit complete, subset of the 
most general eigenfunctions of the CS model. Making use of the following
identity for the Hermite polynomials$^{12}$
\begin{equation}
{ \left(\sum_{k=1}^r {\beta}_k^2\right)^{I/2}\over  I!}   H_I   \left[ 
{\sum_{k=1}^r {\beta}_k x_k\over
\left(\sum_{k=1}^r{\beta}_k^2\right)^{1/2}} \right]
= \sum_{\sum_{k=1}^r m_k=I}  \prod_{k=1}^r  {{\beta}_k^{m_k}\over  m_k!} 
H_{m_k}(x_k) \qquad ,
\end{equation}
Eq. (14) can be written as
\begin{equation}
\phi_I = \exp(-{1\over2} \sum_i^N x_i^2)\,\,\,   H_I({1\over\sqrt{N}} 
\sum_i^N x_i) \qquad .
\end{equation}
Using the center of mass coordinate $R = {1\over N} \sum_i^N x_i$, the radial 
variable $r = \sqrt{{1\over N} \sum_{i<j}^N (x_i-x_j)^2}$ and the identity
$\sum_i^N x_i^2  =  NR^2 + r^2$, one finds
\begin{equation}
\phi_I  =  \exp(-{1\over2} NR^2)\,\,\,  \exp(-{1\over2}r^2)\,\,\, 
H_I(\sqrt{N} R)\qquad . 
\end{equation}
As shown in Ref. 1, the Hamiltonian $H^\prime$ can be written in 
a separated form as
\begin{equation}
H^\prime = H_R + H_r + r^{-2} H_\Omega \qquad ,
\end{equation}
by making use of $R$, $r$, $(N-2)$ angular coordinates $\Omega_i$ and the 
following identities 
\begin{eqnarray}
\sum_i^N \partial_i^2 & = & {1\over N} {\partial^2\over\partial R^2} 
+     {\partial^2\over\partial      r^2}      +{(N-2)\over      r} 
{\partial\over\partial r} + {1\over{r^2}} \hat{L} \qquad,\\
\sum_{{i,j} \atop {i\ne j}}^N {1\over(x_i-x_j)} \partial_i 
& = & {1\over 2} N(N-1){1\over r}{\partial\over\partial    r} + {1\over   r^2}
\hat{M} \qquad .
\end{eqnarray}
Here, the operators $\hat{L}$ and $\hat{M}$ are identical to the ones given by
Calogero$^1$ and act only on the angular coordinates.
\newpage
Explicitly,
\begin{eqnarray}
H_R & = & -{1\over2N} {\partial^2\over\partial  R^2}  +  {1\over2} NR^2 ,\\
H_r & = & -{1\over2} {\partial^2\over\partial r^2} - {[N-2+(\lambda + \gamma ) 
N(N-1)]\over2} {\partial\over\partial  r}  +  {1\over2} r^2 \qquad, \\
H_\Omega & = & \hat{L} + 2(\lambda + \gamma ) \hat{M} \qquad .
\end{eqnarray}
As is clear from above, $H_R$ describes the center of mass degree of freedom.

It has been further established in Ref. 1 that, with a class of symmetrical 
homogeneous polynomials $P_k(x)$ of degree $k$, the function $r^{-k}P_k(x)$ is 
an eigenfunction of $H_{\Omega}$ with the eigenvalue
$E_\Omega =  -k[k+N-3+N(N-1)(\lambda + \gamma )]$.
These eigenfunctions depend only on the $N-2$ angular coordinates $\Omega_i$ 
and are independent of both the radial and center of mass coordinates. The 
eigenfunctions for $H_R$ and $(H_r +  r^{-2} E_\Omega)$ are respectively given
by,
\begin{equation}
\chi_I(R)  =  \exp\left(-{1\over2} NR^2\right) H_I(\sqrt{N}R)  \qquad,\\
\end{equation}
\mbox{and}
\begin{equation}
\eta_{n,k}(r)  = r^k  \exp\left(-{1\over2} r^2\right) L_n^{b+k}(r^2) \qquad . 
\end{equation}
Here, $b={1\over2}(N-3)+{1\over2} N(N-1)(\lambda + \gamma )$.

The total wavefunction for Eq. (19) is
\begin{equation}
\phi_{I,n,k}  =  \exp\left(-{1\over2}  NR^2\right) \,\,\,H_I(\sqrt{N} R)\,\,\,
\exp\left(-{1\over2} r^2\right)\,\,\,  L_n^{b+k}(r^2)\,\,\,   P_k(x) 
\end{equation}
and the corresponding energy eigenvalue is given by
\begin{equation}
E_{I,n,k} = I + 2n+k+E_o  \qquad .\\
\end{equation}
For  $n=k=0$, Eq. (27) becomes
\begin{equation}
\phi_{I,o,o} = \exp\left(-{1\over2}NR^2\right)\,\,\, \exp\left(-{1\over2} 
r^2\right)\,\,\, H_I(\sqrt{N}R) \qquad .
\end{equation}
which is exactly the same as the one given by VOK.

It is worth pointing out that the integers $n$ and $k$ originate from the
radial and angular degrees of freedom respectively. For the special case of the
Calogero Hamiltonian,$^1$ where the harmonic potential is of the type 
$V = {1\over{2N}} \sum_{i<j} (x_i-x_j)^2$, the center of mass motion can be 
eliminated completely. In that case, as is well known, the energy eigenvalues 
are $E_{n,k} =  2n+k+E_o$. The translation invariant wavefunctions for the 
above case can be obtained from Eq. (27) by putting $R=0$. In the CS case, the
presence of the center of mass motion brings in the quantum number $I$.

In conclusion, although the basis set given by VOK forms a complete set, it
only describes the center of mass motion. The present method can also be 
extended to the wavefunctions given by the same authors for the spin
Hamiltonians with identical result. It is of great interest to generate the 
full set of degenerate eigenfunctions by generalizing the algebraic procedure 
advocated here.$^{13}$ It is also of interest to consider the algebra of the 
operators which will connect the degenerate wavefunctions, since the 
degeneracies have a deep connection with symmetries. It should be noted in this
context that, recently various infinite dimensional algebras, {\it e.g.}, the 
$W_{\infty}$-algebra, have manifested in the CS model.$^{14}$ Construction and
analysis of the coherent states$^{15}$ is another direction worth exploring, 
since they will provide a better understanding of the semi-classical behavior 
of this quantum mechanical model. Whether these algebraic techniques can be 
applied to other $N$-body Hamiltonians$^{16}$ is also an important question. 
These problems are currently under investigation and will be reported 
elsewhere. 

We acknowledge useful discussions with Profs. V. Srinivasan, S. Chaturvedi, 
R. MacKenzie and Dr. M. Sivakumar. We also thank Dr. C. Nagaraja Kumar and
R.S. Bhalla for valuable comments. One of us (N.G) would like to thank 
U.G.C (India) for financial support.

\noindent{\bf References}
\begin{enumerate}
\item F. Calogero, {\it Jour. Math. Phys.} {\bf 3}, 419 (1971);\\
B. Sutherland, {\it Jour. Math. Phys.} {\bf 12}, 246 (1971).
\newpage
\item F.D.M. Haldane, {\it Phys. Rev. Lett.} {\bf 60}, 635 (1988);\\
B.S. Shastry, {\it Phys. Rev. Lett.} {\bf 60}, 639 (1988).
\item For various connections, see the chart in B.D. Simons,
P.A. Lee and B.L. Altshuler, {\it Phys. Rev. Lett.} {\bf 72}, 64 (1994).
\item See M.L. Mehta, {\it Random Matrices}, Revised Edition (Academic Press 
N.Y, 1990).
\item A.M. Perelomov, {\it Theor. Math. Phys.} {\bf 6}, 263 (1971);\\
M.A. Olshanetsky and A.M. Perelomov, {\it Phys. Rep.} {\bf 71}, 314 (1981); 
{\bf 94}, 6 (1983). 
\item M. Vasiliev, {\it Int. J. Mod. Phys.} {\bf A 6}, 1115 (1991);
A.P. Polychronakos, {\it Phys. Rev. Lett.} {\bf 69}, 703 (1992);
L. Brink, T.H. Hansson and M. Vasiliev, {\it Phys. Lett.} {\bf B 286},
109  (1992); L. Brink, T.H. Hansson, S.E. Konstein and M. Vasiliev,
{\it Nucl. Phys.} {\bf B 384}, 591 (1993).
\item F. Calogero, {\it Jour. Math. Phys.} {\bf 10}, 2191 (1969);\\
P.J. Gambardella, {\it J. Math. Phys.} {\bf 16}, 1172 (1975).
\item K. Vacek, A. Okiji and N. Kawakami, {\it J. Phys. A: Math. Gen.} 
{\bf29}, L201 (1994).
\item F. Lesage, V. Pasquier and D. Serban {\it Nucl. Phys.} {\bf B 435},
585 (1995);\\
Z.N.C. Ha, {\it Nucl. Phys.} {\bf B 435}, 604 (1995).
\item F.D.M. Haldane, {\it Phys. Rev. Lett.} {\bf 67}, 937 (1988);\\
M.V.N. Murthy and R. Shankar, {\it Phys. Rev. Lett.} {\bf 73}, 3331 (1994).
\item C.F. Dunkl, {\it Amer. Math. Soc.} {\bf 311}, 167 (1989).   
\item I.S. Gradshteyn and I.M. Ryzhik, {\it Tables of Integrals, Series
and Products} (Academic Press Inc., 1965).
\item For the CS model on a circle, the Jack polynomials have been recently
obtained algebraically: L. Lapointe and L. Vinet, {\it Exact operator solution
of the Calogero-Sutherland model}, Centre de math\'{e}matiques
preprint, CRM-{\bf 2272} (1995).
\item K.H. Hikami and M. Wadati, {\it Jour. Phys. Soc. Jpn.} {\bf 62}, 4203
(1993); {\it Phys. Rev. Lett.} {\bf 73}, 1191 (1994);\\
H. Ujino and M. Wadati, {\it Jour. Phys. Soc. Jpn.} {\bf 63}, 3385 (1994);\\
H. Ujino and M. Wadati, {\it Jour. Phys. Soc. Jpn.} {\bf 64}, 39 (1995);\\
V. Narayanan and M. Sivakumar, preprint hep-th/9510239;\\
S.B. Isakov and J.M. Leinaas, preprint hep-th/9510184.
\item G.S. Agarwal and S. Chaturvedi, {\it J. Phys.} {\bf A 28}, 5747 (1995).
\item A. Khare, preprint hep-th/9510096.
\end{enumerate}  
\end{document}